\documentclass[12pt]{article}
\usepackage{times}
\usepackage{geometry}
\usepackage[utf8]{inputenc}
\usepackage{enumitem,amssymb}
\usepackage{ragged2e}
\usepackage{fancyhdr}
\usepackage{graphics,graphicx}
\usepackage{color}
\usepackage{xcolor}
\usepackage{lineno}
\usepackage{geometry}
\usepackage{wasysym}

\usepackage[sort&compress,numbers]{natbib}

\usepackage[compact]{titlesec}

\newlist{thematic}{itemize}{8}
\setlist[thematic]{label=$\square$}
\usepackage{pifont}
\newcommand{\cmark}{\ding{51}}%
\newcommand{\done}{\rlap{$\square$}{\raisebox{2pt}{\large\hspace{1pt}\cmark}}%
\hspace{-2.5pt}}

\begin{document}
\raggedright
\huge
Astro2020 Science White Paper \linebreak

A Summary of Multimessenger Science with Neutron Star Mergers \linebreak
\normalsize

\noindent \textbf{Thematic Areas:} \hspace*{60pt} $\square$ Planetary Systems \hspace*{10pt} $\square$ Star and Planet Formation \hspace*{20pt}\linebreak
$\done$ Formation and Evolution of Compact Objects \hspace*{31pt} $\done$ Cosmology and Fundamental Physics \linebreak
  $\square$  Stars and Stellar Evolution \hspace*{1pt} $\square$ Resolved Stellar Populations and their Environments \hspace*{40pt} \linebreak
  $\square$    Galaxy Evolution   \hspace*{45pt} $\done$             Multi-Messenger Astronomy and Astrophysics \hspace*{65pt} \linebreak
  
\textbf{Principal Author:}

Name:	Eric Burns
 \linebreak						
Institution:  NASA Goddard
 \linebreak
Email: eric.burns@nasa.gov
 \linebreak
Phone:  +1-301-286-4664
 \linebreak

\textbf{Co-authors:} 

A. Tohuvavohu (Pennsylvania State University), J. Buckley (Washington University), T. Dal Canton (NASA Goddard), S. B. Cenko (NASA Goddard), J. W. Conklin (University of Florida), F. D'Ammando (INAF - Istituto di Radioastronomia), D. Eichler (Ben-Gurion University), C. Fryer (Los Alamos National Laboratory), A. J. van der Horst (George Washington University), M. Kamionkowski (Johns Hopkins), M. Kasliwal (Caltech), R. Margutti (Northwestern University), B. D. Metzger (Columbia University), K. Murase (Pennsylvania State University), S. Nissanke (GRAPPA), D. Radice (Princeton), J. Tomsick (UC Berkeley),  C. A. Wilson-Hodge (NASA Marshall), B. Zhang (University of Nevada, Las Vegas)
  \linebreak
  
\justify
\textbf{Abstract:}

Neutron star mergers, referring to both binary neutron star and neutron star black hole mergers, are the canonical multimessenger events. They have been detected across the electromagnetic spectrum, have recently been detected in gravitational waves, and are likely to produce neutrinos over several decades in energy. The non-thermal prompt and afterglow emission of short gamma-ray bursts and the quasi-thermal emission from the radioactively powered kilonovae provide distinct insights into the physics of neutron star mergers. When combined with direct information on coalescence from gravitational waves and neutrinos these sources may become the best understood astrophysical transients. Multimessenger observations of these cataclysmic events will determine sources of gravitational waves and astrophysical neutrinos, enable precision cosmology, and unique tests of fundamental physics, the origin of heavy elements, the behavior of relativistic jets, and the equation of state of supranuclear matter. In this white paper we present a summary of the science discoveries possible with multimessenger observations of neutron star mergers and provide recommendations to enable them in the new era of time-domain, multimessenger astronomy.

\pagebreak

\section{Introduction}
The first multimessenger detection of a neutron star (NS) merger was jointly detected in gravitational waves (GWs) and $\gamma$-rays \citep{GW170817_LVC,GW170817_GBM,GW170817_INTEGRAL,GW170817-GRB170817A}. The aftermath was detected in follow-up observations across the electromagnetic (EM) spectrum \citep[e.g][]{GW170817_afterglow_Xray_discovery,GW170817_afterglow_discovery_radio,GW170817_afterglow_optical_discovery, xray_afterglow_fading,GW170817_kilonova_optical,GW170817_kilonova_UV,GW170817_kilonova_infrared}. We describe the outstanding science possible with observations of NS mergers, followed by the capabilities necessary for discovery. 

\section{Science}
This science is predicated on astrophysical observations of these phenomena; therefore, we first discuss the astrophysics of NS mergers and follow with the broader science they enable. 

\subsection{Astrophysics}
The detections and associations of GW170817, GRB 170817A and the kilonova, AT2017gfo, confirmed decades old predictions from theory, observation, and simulation on the nature of short gamma-ray bursts (SGRBs) and the origin of heavy (r-process) elements \citep[e.g.][]{NSBH_rprocess_Lattimer, Blinnikov1984, Eichler_1989, Paczynski1991, Narayan1992, jets_neutrino_antineutrino, NSM_Rosswog_1998, Li_1998, NSM_Bloom_1999, Popham_1999, Fryer_1999, SGRBs_fong}. We first discuss the (mostly) distinct SGRB and kilonovae transients, followed by astrophysical highlights informed by studies of these sources. 

\subsubsection{Short Gamma-Ray Bursts}
Gamma-Ray Bursts (GRBs) originate from collimated, relativistic outflows. Their prompt emission are the most luminous (EM) events in the universe, but we do not understand where or how the $\gamma$-rays are produced \citep[see, e.g.][for reviews]{GRBs_synchrotron,GRBs_photospheric}. Prompt emission is followed by afterglow from synchrotron radiation as the jet interacts with surrounding material \citep{GRB_afterglow_synchrotron}, observations of which enable studies of the local environment and energetic efficiencies \citep[e.g.][]{SGRBs_fong}. GRB 170817A was underluminous compared to the existing sample of SGRBs, and the temporal behavior of its afterglow had never been previously observed \citep[e.g.][]{GW170817_GBM,GW170817-GRB170817A,GW170817_afterglow_Xray_discovery,GW170817_structured_jet_margutti}. This led to significant debate on the structure of the non-thermal emitting region \citep[e.g.][]{GW170817_kilonova_UV,GW170817_afterglow_Xray_discovery,GW170817_structured_jet_margutti, GW170817_cocoon_Mooley,GRB170817A_Eichler,GW170817_afterglow_latettime_2018,GW170817_VLBI_Mooley,ghirlanda2019compact} and the origin of the prompt emission \citep[e.g.][]{GW170817_cocoon_Kasliwal,GW170817-GRB170817A,GRB170817A_prompt_example,KM_1}.

Multimessenger studies of NS mergers will determine the origin and implications of their non-thermal emission (prompt, afterglow, and more) \citep[e.g.][]{GW170817-GRB170817A,GRB170817A_prompt_example,SGRB_precursors,SGRB_EE,SGRB_magnetar_rowlinson_2013,SGRB_precursors,tsang2012resonant,nonthermal_Murase} and their relation to the different possible central engines \citep[e.g.][]{SGRB_magnetar_lu_2015, SGRB_magnetar_metzger_2008}, the structure of SGRB jets and their interaction with the kilonova material \citep[e.g.][]{GW170817_kilonova_UV}, and how they relate to the intrinsic parameters of the progenitor system. These studies may soon be augmented by TeV detections of SGRBs \citep{TeV_GRB_1,TeV_GRB_2,CTA_1,CTA_2,TeV_add_2,TeV_add_3}. Well characterized events will uncover the fraction of SGRBs from binary neutron star (BNS) and neutron star black hole (NSBH) mergers, and the differences between the two. The merger time from GWs and the onset time of $\gamma$-rays in joint GW-GRB detections will constrain the bulk Lorentz factor of the jet (which determines the observable region of the jet due to Doppler beaming) and the size of the $\gamma$-ray emitting region \citep[as in][]{GW170817-GRB170817A}, and provide new information to uncover the origin of the prompt emission \citep[e.g.][]{GW170817-GRB170817A,GRB170817A_prompt_example}. 

\subsubsection{Kilonovae}
Prior searches for kilonovae following well-localized SGRBs resulted in a few claims of detection \citep{KN_130603B,KN_060614,KN_050709,KN_magnetar,KN_150101B} and interesting non-detections \citep[e.g.][]{kilonova_diversity_observed}; the only unambiguous, well-studied kilonova is the one that followed GW170817: AT2017gfo. This event met broad expectations, and a general picture has emerged: a bright blue kilonova from the relatively lanthanide-poor material in the polar regions followed by a transition to a fainter, progressively redder kilonova from lanthanide-rich material in the planar region \citep[e.g.][]{GW170817_kilonova_theory}. However, the  ultraviolet emission at 12 hours post merger was brighter than some expectations \citep[e.g.][]{metzger2010electromagnetic,GW170817_kilonova_UV} and earlier optical and ultraviolet observations are key to determining the origin of this emission \citep{GW170817_kilonova_models_early}. 

Detections of kilonovae following GW triggers will reveal the relationship between kilonova color and luminosity to properties of the progenitor, including merger type (BNS or NSBH), the short-term remnant object (BH, metastable NS, stable NS), the mass and spin of the progenitors, constrain the NS equation of state (EOS), and ultimately tie the observed diversity to specific physical processes \citep[see][for a review]{review_kilonovae_metzger}.

\subsubsection{Neutron Star-Black Hole Systems}
NSBH mergers are also predicted to result in both SGRBs and kilonova, so long as the NS is disrupted outside of the innermost stable circular orbit. However, there is no known stellar system that is definitely a NSBH binary. Since NSBH mergers have larger detectable volumes than BNS mergers \citep[e.g.][]{LVCK_prospects} we should expect a detection in the next few years, unless they are significantly less common. The first classification of a NSBH merger will likely require multimessenger observations (unless the GW signal is particularly loud). The BH can be inferred from the GW mass measures and the NS from an associated SGRB or predominantly red kilonova. Such an observation would prove NSBH systems exist by observing one merge.

\subsubsection{Stellar Formation and Evolution}
Population synthesis studies bring together models of stellar formation, stellar evolution, binary evolution, and supernovae to make predictions on the properties of compact binaries. The formation channels of compact binary systems that merge in a Hubble time include the evolution of field binaries \citep[see][for a review]{BNS_review_rasio, dominik_2013} or dynamical capture. Multimessenger observations can classify GW triggers as NS mergers, measure mass and spin distributions, the offset from their host galaxies, the local volumetric merger rates of these systems and the source evolution of those rates. These observations with population synthesis studies can differentiate between formation channels, constrain the fraction of stellar systems that result in NS mergers \citep[see, e.g.][for results from GW170817]{GW170817_LVC_progenitor}, the initial offset distribution for the compact objects, the fraction of supernovae that result in NS or BH remnants and this relation to progenitor mass \citep[see, e.g.][]{merger_rates_review_Sadowski}. 

\subsection{Cosmology}
$H_0$ sets the age and size of the universe. It is one of six parameters in the base $\Lambda$CDM concordance cosmological model. It can be measured directly in the nearby (late) universe from type Ia supernovae studies \citep[e.g.][]{H0-Riess_2018}, or inferred from observations of the distant (early) universe from studies using observations of the Cosmic Microwave Background (CMB), Baryon Acoustic Oscillations (BAO), and an assumed cosmological model \citep[e.g.][]{Planck_2018_cosmo}. Comparing $H_0$ values from observations of the early universe against those from observations of the late universe provides a stringent test of cosmological models. These measures disagree with 3.8$\sigma$ significance \citep{Riess_H0_38} with no obvious systematic origin \citep{H0_Riess_2016,Planck_2018_cosmo}. 

Mergers of compact objects are standard sirens with an intrinsic luminosity predicted from GR which can be combined with an associated redshift to measure $H_0$ \citep{H0-Schutz-1986,schutz_2002}. The precision is limited by the correlated inclination and distance determination from GW measures, which can be ameliorated for events with an associated SGRB \citep[e.g.][]{dalal_2006,guidorzi_2017_H0,GW170817_VLBI_Mooley,hotokezaka_2018_H0}. Over the next decade standard siren cosmology could resolve the current $H_0$ controversy and independently calibrate the cosmological ladder \citep[see][for a review]{freedman2001final}. Further, as the GW interferometers improve and detect NS mergers deep into the universe, those with (EM-determined) redshift will create the most precise Hubble Diagram spanning from the local universe to deep into the deceleration era \citep{GW_IFO_3rd_Gen_EU}. Combining standard siren cosmology with CMB+BAO observations could resolve the neutrino mass hierarchy question and improve constraints on the effective number of neutrino species and the EOS of dark energy \citep{Planck_2018_cosmo}. Further adding information from forthcoming transition era cosmology experiments (e.g. LSST, EUCLID, WFIRST) gives sub-percent precision cosmology throughout the universe, allowing studies on multi-parameter extensions to $\Lambda$CDM (e.g. $\Omega_k$, $w_0$, and $w_a$ simultaneously). 

\subsection{Fundamental Physics}
The $\sim$seconds difference in the arrival times of gravity and light across cosmological baselines enable observations of NS mergers to probe some fundamental aspects of physics far greater than any other method. Detections with GWs and GRBs in $\sim$keV-MeV energy (which do not undergo significant extinction, dispersion, nor absorption) provide the best measure of the speed of gravity \citep{GW170817-GRB170817A}, and relative violations of the Weak Equivalence Principle \citep[e.g.][]{Review_Tests_of_GR_overall_Will_2014,GW170817-GRB170817A, WEP_2,KM_1} and Lorentz Invariance (LIV) \citep{LIV_SME,GW170817-GRB170817A,KM_1}. GeV-TeV (EM) observations of SGRBs have the greatest discovery space, and current best constraints, on general LIV \citep{LIV_090510_limit_2009}. These observations test the Special Theory of Relativity which has been woven into all of modern physics, the Einstein Equivalence Principle that all metric theories of gravity must obey, the quantization of spacetime itself (and therefore quantum gravity), and measure fundamental constants of the universe \citep[see e.g.][]{2015CQGra..32x3001B}. 

\subsection{The Origin of Heavy Elements}
For many decades, astronomers have debated whether NS mergers or Core-Collapse Supernovae (CCSN) are the site of heavy (above the iron peak), rapid neutron capture (r-process) elements \citep[see][for a summary of the arguments]{qian_2000_summary}. Even with the well-sampled AT2017gfo we are unsure if all three abundance peaks were synthesized. With a greater understanding of the r-process production in NS mergers from future kilonovae observations and improved measures on the rates of such events from GW observations, we can determine the total and relative heavy elemental abundances produced from these events. This can be compared to the observed solar system abundances to determine the fraction of heavy elements that come from NS mergers, as opposed to CCSN \citep[see e.g.][]{rprocess_1_siegel, rprocess_2_cote, rprocess_3_drout}. CCSN are thought to track the cosmic stellar formation rate, as are BNS and NSBH mergers modulo their inspiral times. An improved understanding of the source evolution of NS mergers will determine the heavy element abundance through cosmic time.

\subsection{Relativistic Jets and Particle Acceleration}
Many astrophysical sources, such as blazars, microquasars, and protostars, arise from collimated outflows referred to as jets; GRB jets are an ultrarelativistic version. Multimessenger observations provide direct measures of the central engine, which when combined with EM observations of the SGRB will uncover key information about jets. We will learn if SGRB jet formation requires an event horizon \citep{EH_Eichler} (i.e. a BH central engine), or if they can be created by magnetar central engines \citep[see, e.g.][]{review_grb_central_engine,TeV_GRB_3}. Multimessenger observations can determine required energetics efficiencies that may delineate between models of jet formation \citep[e.g.][]{jets_blandford_znajek,jets_neutrino_antineutrino,jet_process_gw170817, jet_form_2, TeV_GRB_1}. The structure of GRB jets can be studied from observations of the non-thermal emission, and move our understanding beyond the on-axis top-hat jet models rejected by GRB 170817A. Relativistic jets may be either hadron or magnetically dominated, which could be determined from (non-)detections of neutrinos \citep[e.g.][]{Zhang_Kumar,TeV_GRB_1,TeV_GRB_2,KM_2,KM_3,UHECR_A} with sufficiently sensitive detectors. The origin of cosmic rays beyond the knee or second knee energies is unknown, and neutron star mergers have been suggested among the promising candidate sources \citep{UHECR_A, UHECR_B}.

\subsection{Neutron Star Equation of State}
NSs are the densest known matter in the universe, far beyond anything achievable in terrestrial laboratories. The NS EOS characterizes the density-pressure relationship \citep[see][for a review]{NS_EOS_review_Lattimer}. From an assumed EOS, several observables can be predicted, such as a mass-radius relation \citep[e.g.][]{NS_EOS_review_Ozel}. In BNS mergers the remnant object just after merger can be a stable NS, a metastable NS, or a BH, with the NS EOS determining which case for a given merger. These options can be resolved by observations of associated SGRBs or kilonovae with specific characteristics \citep[e.g.][]{SGRB_EE, SGRB_magnetar_metzger_2008, SGRB_magnetar_rowlinson_2013, metzger2014optical, siegel2016electromagnetic, nonthermal_Murase, GW170817-GRB170817A,NS_EOS_metzger,NS_EOS_Radice_2017, NS_EOS_margalit_2017, NS_EOS_radice_2018} and eventually directly confirmed with GW or MeV neutrino observations \citep{MeV_neutrino_1,MeV_neutrino_2}. Multimessenger observations of NSBH mergers provide additional constraints as the orbital radius for NS disruption is a strong function of the mass and spin of the BH and the NS radius \citep{NS_EOS_foucart_2012}. Multimessenger observations of NS mergers will provide information, complementary to EM-only or GW-only constraints, to determine the EOS of supranuclear matter, and possibly constrain the phase diagram of quantum chromodynamics \citep[e.g.][]{NS_EOS_most_2018, NS_EOS_bauswein_2018}. 

\section{Recommendations}
NS mergers emit across decades in energy in several messengers which enables greater understanding but requires vast observational resources. Prompt $\gamma$-ray, afterglow (TeV to radio), and TeV-PeV neutrino observations will provide new insight into SGRBs, relativistic jets, and particle acceleration. Ultraviolet, optical, and infrared (UVOIR) observations uncover and characterize kilonovae and the origin of heavy elements. Probing fundamental physics requires GW and prompt GRB detections. Cosmological studies require GW distance and EM redshift determinations. All observations help constrain the NS EOS. Much of this science is only possible with a population of events. Information on the inspiral, coalescence, or jet before the coasting phase (obtained from GW, $\gamma$-ray and neutrino detectors) can only be done from serendipitous observations, requiring all-sky monitors with high duty cycles and large fields of view.

Some detections also enable other observations. Despite collimation, joint GW-GRB detections are expected to be reasonably common because of preferential selection effects \citep[combining information from][]{GW_inclination,SGRBs_fong} and the increase of GW detections from GRB observations \citep[e.g.][]{joint_williamson, joint_blackburn}. Joint GW-GRB detections further constrain the region of interest for follow-up. Arcsecond localizations are necessary for host-galaxy identification and redshift determination and are only possible from follow-up observations. The earliest follow-up signal that can be detected is the GRB afterglow; however, based on the prompt vs afterglow brightness of GRB 170817A, we do not expect off-axis afterglow detections (with no detectable prompt emission) to be common. When there is no detectable SGRB, the first EM emission for a blue kilonova appears to be in the UV and in optical or infrared for red kilonovae (on longer timescales).

Kilonovae will be the dominant EM counterpart for nearby NS mergers as they are omnidirectional, but SGRBs will dominate for distant events as they can be significantly brighter. While we split our recommendations between the next decade and beyond, regardless of time period, \textit{we recommend vigorous funding for upgraded GW interferometers}.


\subsection{The Next Ten Years}
With the full design network of current GW detectors, localizations sufficient for existing optical facilities will be relatively common \citep{LVCK_prospects}. The funded LIGO A+ upgrade (nominally available in 2024), will detect a NS merger roughly once a week. Follow-up ground observations with current or expected missions (e.g. HAWC, CTA, VLA, SKA, ZTF, LSST) reliably cover TeV, optical, near-IR, and radio. As kilonovae transition to redder emission over time, the IR-only emission is sufficiently late ($\sim$days) that these wavelengths can be covered by JWST and WFIRST observations. \textit{We endorse the allocation of appropriate observing time and target of opportunity programs for pointed telescopes and directly recommend LSST follow-up of NS mergers}. Megaton-class MeV neutrino detectors could detect particularly nearby NS mergers. Initial upgrades to TeV-PeV neutrino detectors are on-going; \textit{we endorse the full IceCube Gen-2} \citep{IceCube-Gen2}. 

The critical wavelengths detectable only in space are keV-MeV $\gamma$-rays, X-rays, and UV. As such, we recommend the extension of the \textit{Fermi} mission (because it detects more prompt SGRBs than all other active mission combined) and the \textit{Neil Gehrels Swift Observatory} (primarily for fast-response X-ray and UV coverage). To capture the full range of possible kilonova colors, \textit{we recommend wide-field UV (space-based) and NIR (ground-based) facilities}. To enable time-domain astronomy, \textit{we recommend improvements to real-time communication for space-based missions.}

Beyond the required observational capabilities, \textit{we strongly recommend greater NSF-NASA cooperation} as both agencies have assets critical for multimessenger science. \textit{Specific recommendations include the funding of beam studies to understand the nuclear processes that currently limit kilonova models \citep[e.g.][]{review_kilonovae_metzger, FRIB_rprocess}, robust funding of multimessenger simulation and theory studies via the TCAN program (created in response to Astro2010), and the consideration of resolving grand problems through NASA-NSF partnership (akin to the DRIVE initiative from the 2013 Solar and Space Science Decadal).} \textit{Technical improvements include improvements to real-time reporting, automated multimission and multimessenger searches, and prompt reporting of initial parameter estimation from GW detections (e.g. masses) to enable follow-up prioritization.}

\subsection{Future Large-scale Missions}
The proposed LIGO-Voyager upgrade has a nominal timeline of $\sim$2030 and it would detect several NS mergers per week with some at cosmological distances ($z>0.1$). \textit{Proposed large-scale missions in the Astro2020 Decadal must be designed for the 2030s, not the current era.} Proposed third generation GW interferometers \citep{GW_IFO_3rd_Gen_US,GW_IFO_3rd_Gen_EU} with best-case timelines of mid-to-late 2030s could detect dozens of NS mergers per day with some beyond the transition era. Fewer upgraded interferometers will result in poorer GW localizations for distant events. To enable all NS merger science \textit{we recommend future GW interferometers aim for broadband sensitivity improvements}. With these upgrades a significant fraction of SGRBs will have associated GW emission. Therefore, \textit{we recommend a large-scale $\gamma$-ray observatory that will detect far more SGRBs than any prior mission (through improved $\sim$keV-MeV sensitivity and broad sky coverage) and localize them to sufficient accuracy for sensitive follow-up observations.} To enable a full study of these events, including redshift determination, they need to be localized to arcsecond precision. With such a $\gamma$-ray instrument, \textit{we would need well-matched follow-up facilities such as a sensitive, fast-reponse, high spatial resolution X-ray telescope and sensitive UVOIR and radio telescopes.}


\pagebreak

\bibliographystyle{abbrv}
\bibliography{references}

\end{document}